\documentclass[12pt,preprint]{aastex}
\shorttitle{HIGH ENERGY NEUTRINOS IN GAMMA RAY BURSTS}
\shortauthors{ASANO \& NAGATAKI}

\begin{document}

\title{VERY HIGH ENERGY NEUTRINOS ORIGINATING FROM KAONS IN GAMMA-RAY BURSTS}
\author{\scshape K. Asano}
\affil{Division of Theoretical Astronomy, National Astronomical Observatory of Japan,
2-21-1 Osawa, Mitaka, Tokyo 181-8588, Japan}
\email{asano@th.nao.ac.jp}
\and
\author{\scshape S. Nagataki}
\affil{Yukawa Institute for Theoretical Physics, Kyoto University,
Oiwake-cho Kitashirakawa Sakyo-ku, Kyoto 606-8502, Japan}
\email{nagataki@yukawa.kyoto-u.ac.jp}

\date{Submitted; accepted }

\begin{abstract}
We simulate neutrino production in a gamma-ray burst (GRB)
with the most detailed method to date. We show that the
highest energy neutrinos from GRBs mainly come from kaons.
Although there is little chance to detect such neutrinos,
attempts of detection are very important to prove physical conditions in GRBs.
\end{abstract}

\maketitle

\section{INTRODUCTION}

The rapid time variabilities and the compactness problem
(see, e.g., a review by Piran \citep{pir99}) suggest that
gamma-ray bursts (GRBs) should arise from
internal shocks within relativistic flows.
In the standard model, a strong magnetic field
is generated, and electrons are Fermi-accelerated
in shocked regions.
The physical conditions in the shocked region imply
\citep{wax95} that protons may be also Fermi-accelerated
to energies $\sim 10^{20}$ eV.
High-energy protons in the GRB photon field can create high-energy
neutrinos via photopion production 
\citep{wax97,wax99,gue01,der03,asa03,gue04,asa05}.
Future observations of neutrinos will be important to
prove the standard model of GRBs and the particle acceleration theory.
The highest energy of neutrinos brings us information on physical
conditions of GRBs.

Recently, \citet{and05} have shown that the kaon contribution
becomes important for neutrino production from jets in supernovae.
They considered mildly relativistic jets that are much
more baryon-rich than a fireball of GRBs where collisions
among accelerated protons ($pp$) occur efficiently~\citep{razzaque04},
making pions and kaons that decay into neutrinos.
Since high-energy charged mesons will cool down
before they decay into neutrinos,
the highest energy of neutrinos is determined by
the equilibrium of the cooling timescale and the decay timescale of mesons.
Considering synchrotron cooling, the highest
energy of mesons that decay into neutrinos
is proportional to $m^{5/2} t_0^{-1/2}$,
where $m$ and $t_0$ are the mass and
the lifetime of mesons at rest, respectively.
So, \citet{and05} pointed out that
the heavier mass of kaons 
than pions leads to dominance of neutrinos from
kaon decay in the highest energy region.

We note that the contribution of kaons for neutrino production
may also be important even in internal shocks of GRBs. 
In this study, we
calculate the spectrum of neutrinos from GRBs, taking account
of decaying modes of charged kaons into neutrinos.
Also, we consider the contribution of long-lived neutral kaons,
$K^0_{\rm L}$, for neutrino production, which was not taken
into account in the previous work. It is noted that 
$K^0_{\rm L}$ does not cool at all before decay into
neutrinos and has some decaying modes into charged pions
and neutrinos. So, it is expected that the highest energy of
neutrinos comes from the decay of long-lived neutral kaons.

In this Letter, using the Monte Carlo method,
we show that the highest energy neutrinos mainly
come from kaons even in internal shocks of GRBs.
In \S 2, we explain our model and method.
The numerical results are in \S 3.
\S 4 is devoted to discussion.

\section{SIMULATION}

Our method of simulation is essentially
the same as in \citet{asa05} but quantitatively improved.
In this study, we adopt experimental results for the
cross sections of
$p \gamma \to n \pi^+$,
$p \pi^0$, $n \pi^+ \pi^0$,
and $p \pi^+ \pi^-$ \citep{sch03} for
$\epsilon' \leq 2$ GeV,
where $\epsilon'$ is the photon energy in the proton
rest frame.
We neglect the reaction $p \gamma \to p \pi^0 \pi^0$,
because the cross section is too small and
the neutrino production rate is independent of 
this reaction.
Since we do not have experimental data of
multi-pion production for $\epsilon'
\gtrsim 1$ GeV, we extrapolate the cross section
by a constant value.
However, our total photoabsorption cross section
agrees well with the experimental value
for $\epsilon' < 2$ GeV \citep{eid04}.
For the pion production by $n \gamma$,
we adopt the same cross sections as $p \gamma$.
Kaons are produced via
$p \gamma \to \Lambda K^+$, $\Sigma^0 K^+$, and $\Sigma^+ K^0$,
or $n \gamma \to \Lambda K^0$, $\Sigma^- K^+$, and $\Sigma^0 K^0$.
Since there are no detailed and precise data
of kaon production experiments,
we adopt values theoretically obtained for
the cross sections of kaon production from
$p \gamma$ and $n \gamma$ \citep{dre92,lee01}.
As shown in Figure 1, the contribution of
kaon production seems negligible.
However, the importance of kaon production
will be shown later.

\begin{figure}[t]
\centering
\epsscale{1.0}
\plotone{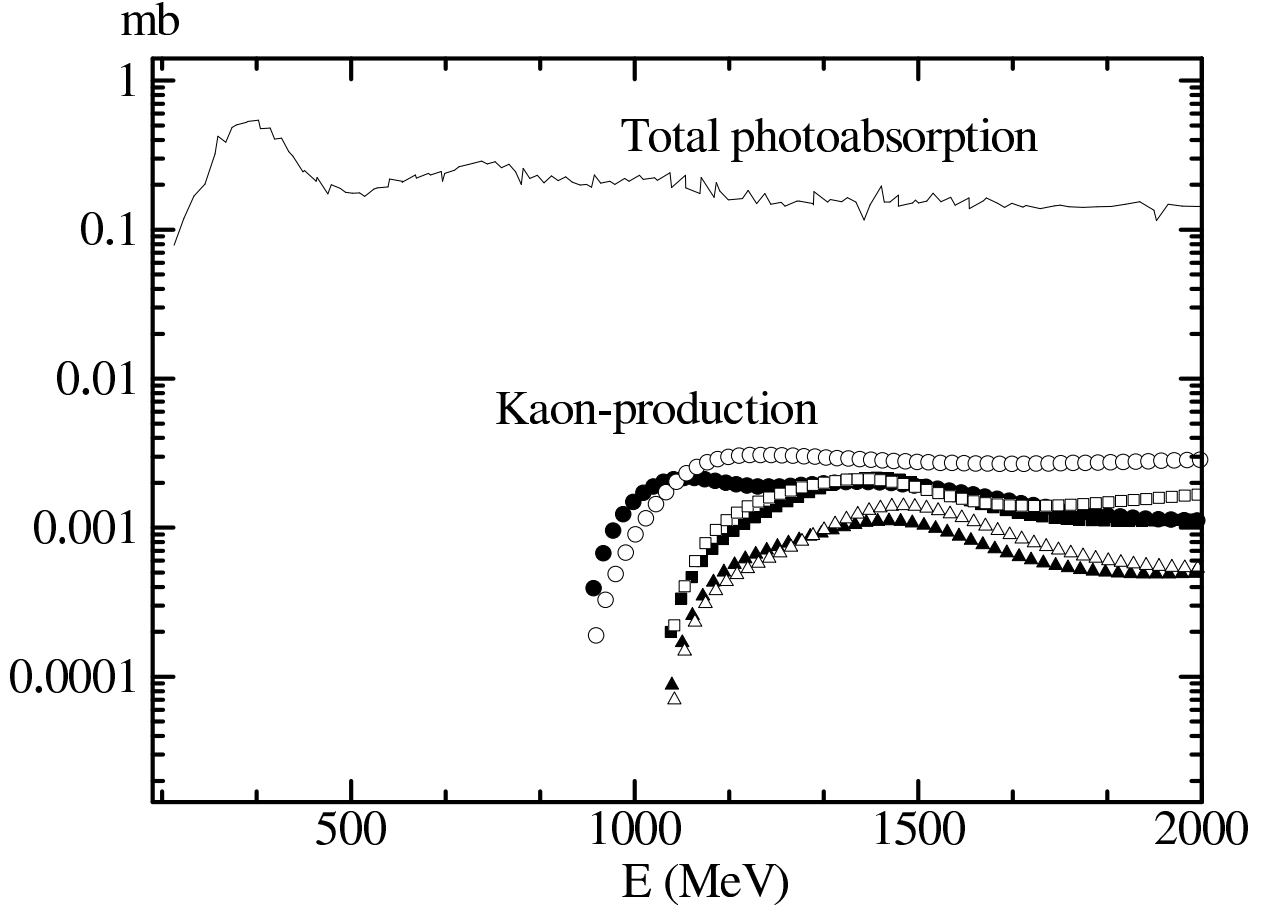}
\caption{Cross sections vs. photon energy in
the baryon rest frame. Solid line is
the total photoabsorption cross section of protons.
Open symbols are for kaon production by $p \gamma$
(circle: $K^+ \Lambda$, square: $K^+ \Sigma^0$,
triangle: $K^0 \Sigma^+$),
while filled symbols are those by $n \gamma$
(circle: $K^0 \Lambda$, square: $K^+ \Sigma^-$,
triangle: $K^0 \Sigma^0$).}
\end{figure}

Of course, $pp$ collisions may also create mesons.
Since the number of target photons is much larger
than the proton number density in our case,
we neglect the effects of $pp$ collisions.

The inelasticity is approximated by a conventional
method as $K=[1-(m_{\rm p}^2-m^2)/s]/2$,
where $s$ is the invariance of the square
of the total four-momentum of the $p \gamma$
($n \gamma$) system,
and $m_{\rm p}$ is the proton mass.
For the double-pion production, we approximate
the inelasticity by replacing $m$ with
$2 m_\pi$.

Our parameter set is similar to that in \citet{der03}:
the total photon energy in a burst
$E_{\rm tot}$, the number of light-curve
pulses (or spikes) $N$,
the Lorentz factor of the shells $\Gamma$,
and shell-collide distances from the central engine $R$.
The number $N$ corresponds to the number
of shells that emit gamma rays.
The photon energy deposited into each shell
is therefore $E_{\rm sh}=E_{\rm tot}/N$.
In the standard model, shells can collide
and emit gamma rays at distances $R$
larger than $3 \times 10^{11} (\Gamma/100)^2 \delta t/(1 {\rm ms})$ cm
from the central sources,
where $\delta t$ is the time between shell ejection events.
For simplification, $R$ and $\Gamma$ are common for all $N$ shells
in this simulation.
The photon number spectrum in the energy range $\epsilon+d \epsilon$
in the shell rest frame
is set at $n(\epsilon) \propto \epsilon^{-1}$
for 1 eV $<\epsilon< $ 1 keV
and $\epsilon^{-2.2}$ for
1 keV $<\epsilon< 10$ MeV.
The break energy 1 keV corresponds to
$100 (\Gamma/100)$ keV in the observer frame.
For $\epsilon<1$ eV, the synchrotron self-absorption may be crucial
\citep{gra00}, while the pair absorption may be crucial for $\epsilon>10$ MeV
(e.g., see Asano \& Takahara 2003, Pe'er \& Waxman 2004).
Although the upper bound of the photon energy
depends on the model parameter because of pair production,
we fix the value as 10 MeV ($\sim 1$GeV in the observer frame).
Since higher energy protons mainly interact with lower energy photons,
the production rate of very high energy neutrinos is not sensitive
to this upper bound.

The shell width in the comoving frame
is assumed to be $R/\Gamma$, as conventionally assumed,
although there is the possibility
of thinner shells \citep{asa02}.
We express the energy density of the magnetic field
as $f_{\rm B}$ times the photon energy density.
In this Letter, we adopt $f_{\rm B}=0.1$.

We inject protons with a number spectrum
proportional to $\epsilon_{\rm p}^{-2}$
above 10 GeV in the shell rest frame.
The maximum proton energy is determined
by the condition that the Larmor radius
is smaller than both the size scale of
the emitting region and the energy-loss
length. We estimate the energy-loss length
using synchrotron, inverse Compton, and photomeson cooling processes.
The total energy of the accelerated protons
in a shell is assumed to be the same as
$E_{\rm sh}$.
Our method pursues energy loss processes of each baryon
via synchrotron, inverse Compton, and photomeson cooling processes
during the dynamical timescale $R/c \Gamma$ in the shell rest frame.

\section{RESULTS}

We have simulated for a wide range of parameters as
$E_{\rm tot}=10^{48}$-$10^{54}$ ergs, $N=10$-$1000$,
$\Gamma=100$-$1000$, and $R=10^{13}$-$10^{15}$ cm.
Of course, larger $E_{\rm tot}$ and smaller $R$
are favorable for neutrino production,
and as \citet{der03} showed, very luminous bursts
are required to detect neutrinos on the Earth.
Therefore, we show only one representative example in this Letter.
The parameter values are
$E_{\rm tot}=10^{54}$ ergs, $N=1000$ ($E_{\rm sh}=10^{51}$ ergs),
$\Gamma=100$, and $R=10^{13}$ cm.
The corresponding variability timescale
$t_{\rm var} \simeq R/\Gamma^2 \sim 30$ ms, which is
not so far from the typical observed timescale
$\gtrsim 0.1/(1+z)$ s.

Since the allowed region of the GRB parameters is wide,
there may be both optically thin and thick sources
to Thomson scattering \citep{mes00}.
Our example would imply that $R$ is close to
the photosphere.
When it is assumed that the energy density of protons is the same as
the photon energy density and the average proton energy
is mildly relativistic ($\lesssim 5 m_{\rm p} c^2$),
the proton number density in the comoving frame
is obtained as $\gtrsim E_{\rm sh}/(20 \pi R^3 m_{\rm p} c^2)$.
This means that the optical depth for the Thomson
scattering is $\sim 1$ for our parameter set.
Photon scatterings do not sufficiently affect the gamma-ray spectrum
for this marginal optical depth.

From our simulation, we obtain spectra of created mesons as is shown
in Figure 2.
One-half of neutral kaons are $K^0_{\rm L}$,
while the rest are $K^0_{\rm S}$.
Since the cross sections of kaon production are
smaller than those of pion production,
the number of kaons is much less than pions.
However, the highest energy charged mesons will cool down
before they decay into neutrinos.

Our results for other parameter sets agree with
the condition of ultra high energy cosmic-ray
production obtained by \citet{asa05}:
$R \gtrsim 10^{14} (E_{\rm sh}/10^{51} {\rm erg})^{1/2}$ cm.
In the case of Figure 2, also protons above $10^{15}$ eV
cool down before they escape from the shell.

\begin{figure}[t]
\centering
\epsscale{1.0}
\plotone{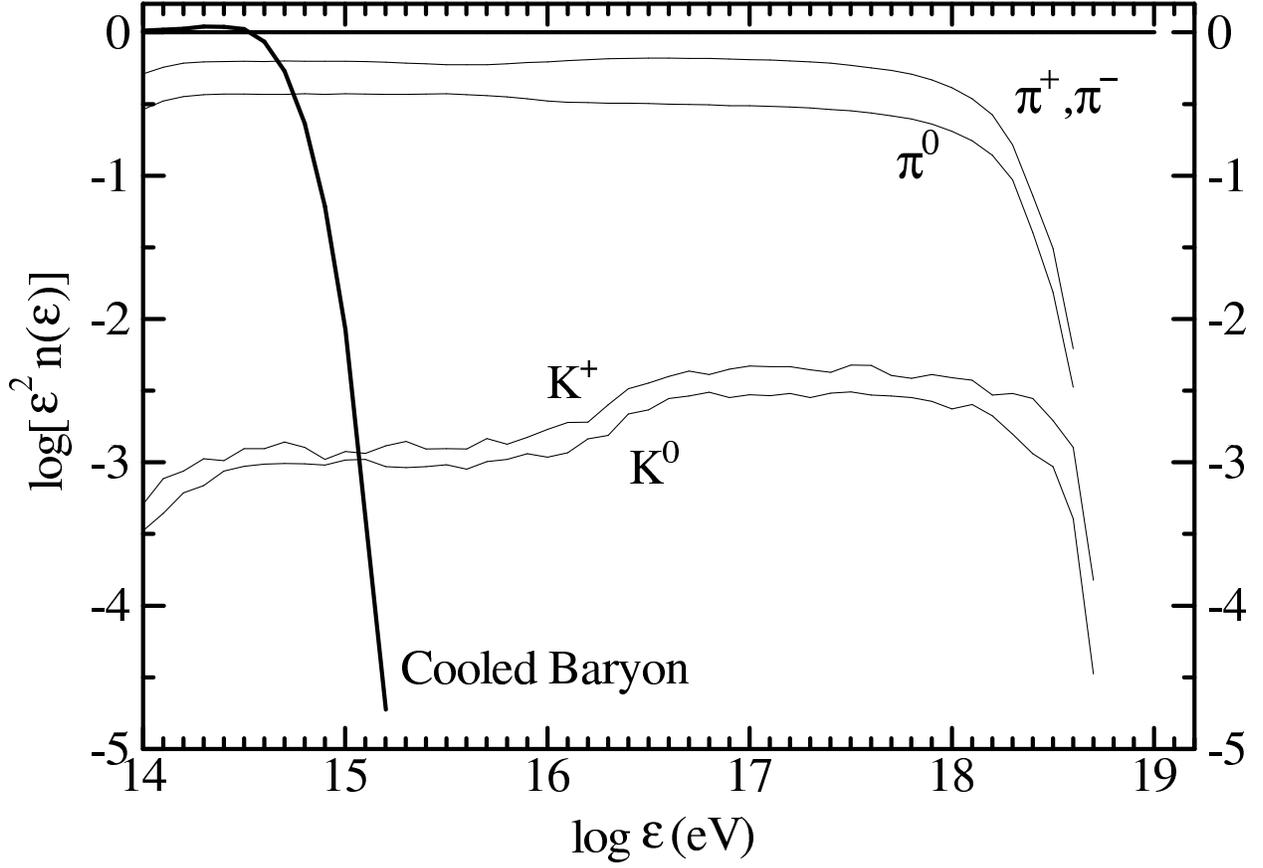}
\caption{Spectra of mesons created by $p \gamma$
and $n \gamma$ reactions in the observer frame.
Values are normalized by the initial proton spectrum.}
\end{figure}

We follow the behavior of
pions and kaons until they decay into positrons (electrons)
and neutrinos using the same method as in \citet{asa05}.
Synchrotron and inverse Compton emissions are
taken into account.
Charged kaons have six decay modes;
$K^+ \to \mu^+ \nu_\mu$ (63\%), $\pi^+ \pi^0$ (21\%),
$\pi^+ \pi^+ \pi^-$ (6\%), $\pi^0 e^+ \nu_e$ (5\%),
$\pi^0 \mu^+ \nu_\mu$ (3\%), and $\pi^+ \pi^0 \pi^0$ (2\%),
while $K^0_{\rm L}$ will decay into
$\pi^+ e^- \bar{\nu_e}$ (39\%), $\pi^+ \mu^- \bar{\nu_\mu}$ (27\%),
$\pi^0 \pi^0 \pi^0$ (21\%), and $\pi^+ \pi^- \pi^0$ (13\%).
In Figure 3, total neutrino spectra emitted from this example are shown.
Although there are fewer kaons than pions, the highest energy neutrinos
originate from kaons around $\epsilon_\nu \sim 10^{18}$ eV.

\begin{figure}[t]
\centering
\epsscale{1.0}
\plotone{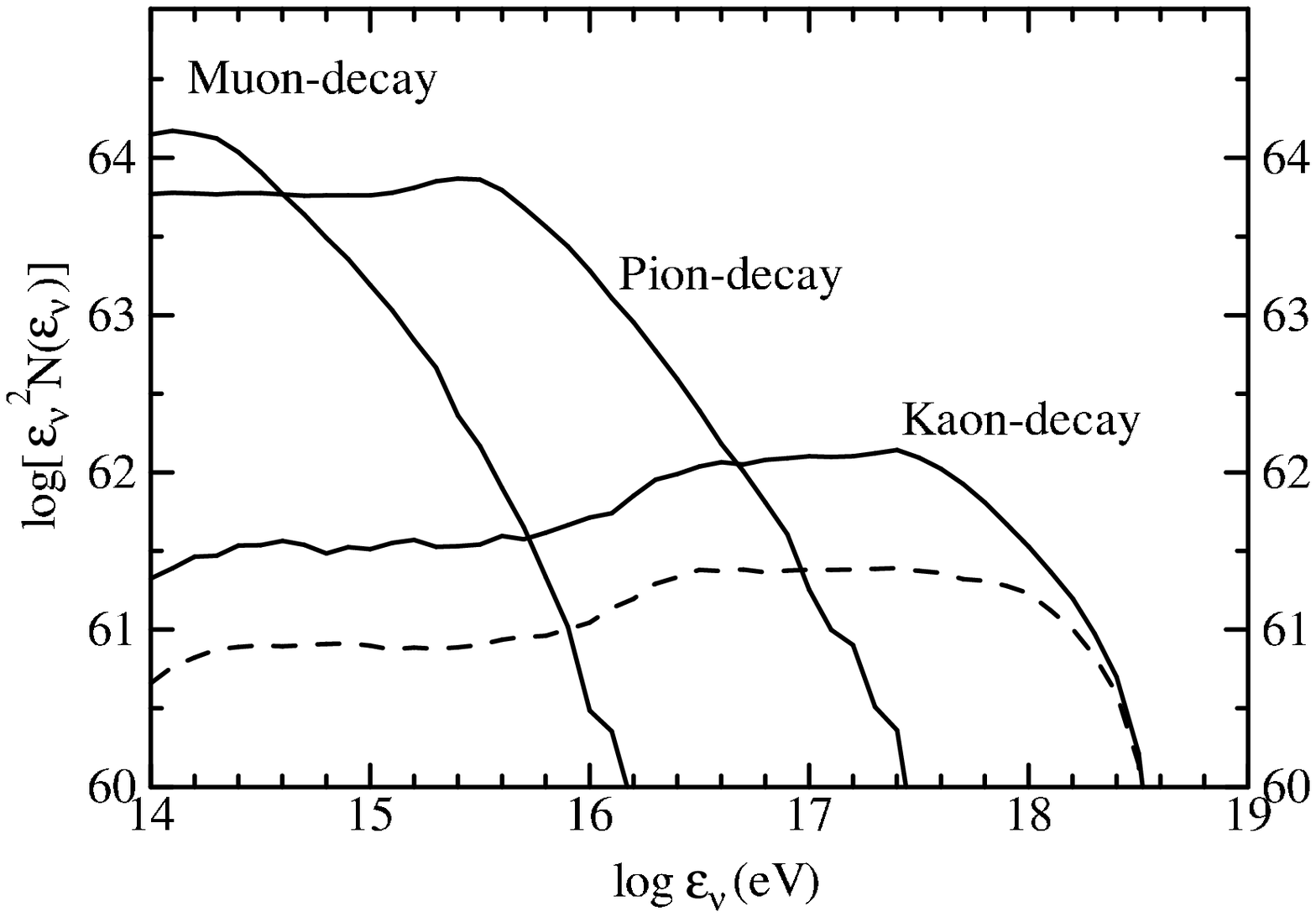}
\caption{Spectra of neutrinos $\epsilon_\nu^2 N(\epsilon_\nu)$
(in units of eV).
Dashed line is for neutrinos from $K^0_{\rm L}$-decay.
}
\end{figure}

Since very high flux is required to detect neutrinos from GRBs
by a $1 {\rm km}^2$ neutrino detector such as IceCube
\citep{der03}, we consider an optimistic case:
a GRB occurs at 30 Mpc, and the detection efficiency
of upward-going neutrinos with energy $\epsilon_\nu$
is assumed to be $10^{-4} (\epsilon_\nu/10^{14} {\rm eV})^{1/2}$
for $\epsilon_\nu >10^{14}$ eV,
although it may be difficult to measure energies of neutrinos
above $10^{17}$ eV by IceCube.
Figure 4 shows the detectable number of spectra of neutrinos
in this case.
The vertical axis $\epsilon_\nu N_{\rm d}(\epsilon_\nu)$
roughly corresponds to the detectable number in each energy range.
In this case, the expectation value of neutrinos from kaons
is 0.1-1 by a $1 {\rm km}^2$ detector.
However, for very high energy neutrinos, it may be possible to build
detectors with effective volume orders of magnitude larger than $1 {\rm km}^2$,
such as the Extreme Universe Space Observatory ($10^5 {\rm km}^2$ detector),
because the Earth is thick for such neutrinos.
From the step-function-like features in the spectra,
we can easily distinguish origins of neutrinos.

\begin{figure}[t]
\centering
\epsscale{1.0}
\plotone{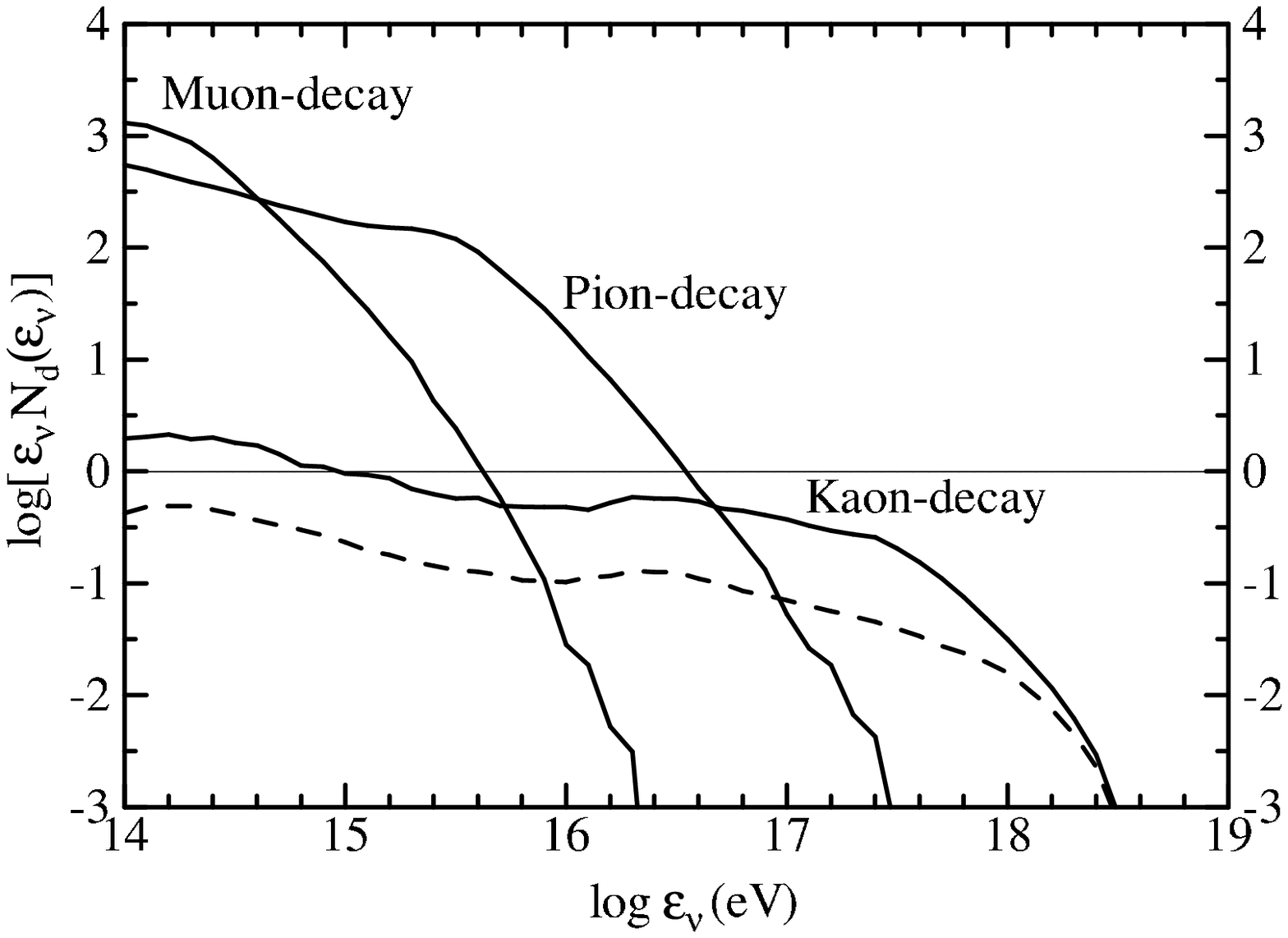}
\caption{Spectra of the detectable number of neutrinos $N_{\rm d}(\epsilon_\nu)$
(in units of neutrinos eV$^{-1}$)
from a GRB of $10^{54}$ ergs at 30 Mpc
by a $1 {\rm km}^2$ detector.
Here $N_{\rm d}(\epsilon_\nu)$ is obtained from
the simple formula of the detection efficiency (see text).
Dashed line is for neutrinos from $K^0_{\rm L}$-decay.
}
\end{figure}

As shown in Figures 3 and 4, $K^0_{\rm L}$-decay neutrinos are
dominant above $10^{18}$ eV, although the flux is too dim
to detect on the Earth.
On the other hand, the contribution of $K^0_{\rm L}$-decay
in the energy band below $10^{18}$ eV is not so prominent.

Since we have considered many decaying modes, the
production ratio of high-energy muon and electron neutrinos is
not 2:1 exactly. However, the neutrinos will be almost equally
distributed between flavors as a result of vacuum neutrino
oscillations~\citep{wax97}. So, there may be a possibility that tau neutrinos
are detected through double-bang events~\citep{athar00}.

\section{DISCUSSION}

As we have shown in this Letter, the highest energy neutrinos
may originate from kaons in GRB internal shocks.
The detection of such neutrinos is very important
to prove the physical conditions in internal shocks
and the particle acceleration theory.
In order to observe such neutrinos,
a bright GRB should occur at a very close distance,
or amazing progress in the technology of neutrino observation
is required.
When we use the observed local GRB rate,
0.4 Gpc$^{-3}$ yr$^{-1}$ \citep{guetta05},
the chance probability of having a source within 30Mpc away
from the Earth is about 1 event per $10^5$ yr.
However, the nearest burst ever observed is GRB 980425 at 40 Mpc,
which was less luminous than usual bursts.
Although a GRB at 30 Mpc seems to be optimistic,
we should prepare for an event beyond expectation,
such as the giant flare from SGR 1806-20 \citep{hur05}.

As a rough standard,
we have used a simple formula of the detection efficiency,
extrapolating the case for upward-going neutrinos.
However, in the very high energy band
downward-going neutrinos are favorable to be detected
rather than upward-going neutrinos,
because the Earth is thick for such neutrinos.
For downward events, the detection efficiency of very high energy neutrinos
would be enhanced.
In addition it may be possible to build
detectors with effective volume orders of magnitude larger than $1 {\rm km}^3$,
which would enhance the chances of detection of the kaon decay neutrinos.

The neutrinos produced from kaon decays may be detected as a neutrino
background, so let us estimate the flux of the neutrino background roughly.
It can be seen from Figure 3 that the fluence of neutrinos $\epsilon_\nu^2
N(\epsilon_\nu)$ at $10^{17-18}$ eV
from this example is $\sim$ 10$^{53}$ GeV.
If we adopt this value,
the resulting neutrino background
level at $10^{17-18}$ eV is $\sim 3 \times 10^{-11}$ GeV cm$^{-2}$
s$^{-1}$ sr$^{-1}$, where we use the local GRB rate,
0.4 Gpc$^{-3}$ yr$^{-1}$, without correction due to jet collimation
\citep{guetta05} and the age of
the universe, $\sim 10^{10}$ yr.
This background level is far below the
detection limit expected from 3 years of IceCube operation.
Since our example is an optimistic case for neutrino emission,
a more realistic background level may become lower 1 or 2 orders
of magnitude than above.
Dominant sources of background neutrinos in the highest energy band
are not sure so far.
If we can observe the background neutrinos at $10^{17-18}$ eV
with a highly efficient detector in the future,
the characteristic flat spectrum predicted from GRB sources
is distinguishable from spectra of other sources,
such as GRB early-afterglows or blazars (see, e.g., Schneider et al. 2002).

\citet{raz04b} showed that
very high energy gamma rays from $\pi^0$-decay
($\gtrsim 10^{17}$ eV) can escape from shells
without electron-positron pair creation.
However, this result largely depends on the assumption
of spectra in the lower energy region.
A recent observation \citep{bla05} shows brighter optical emission
than \citet{raz04b} assumed.
On the other hand, $K^0_{\rm L}$ also has some decaying modes into
neutral pions. So, there is a possibility that delayed
gamma rays come from the decays of $K^0_{\rm L}$. 
We can investigate the validity of this statement by checking
whether $K^0_{\rm L}$ can decay after GRB photons escape
from shells. Since the
mean lifetime of $K^0_{\rm L}$ is 5.17$\times 10^{-8}$ s in the kaon rest frame,
the mean lifetime of $K^0_{\rm L}$, $\Delta t_{\rm k}$, of
energy $\epsilon_{\rm k}$ in the shell rest frame becomes
$\Delta t_{\rm k} = 3.3 (\epsilon_{\rm k} / 10^{16.5} {\rm eV})$ s.
On the other hand, the typical dynamical timescale of GRBs
in the shell rest frame is 3 ($\Gamma/100$)($t_{\rm var}/30 {\rm ms}$) s.
So, we expect that gamma rays produced from kaons with energy
larger than $10^{18.5} (\Gamma/100)$ eV can escape from shells.

Of course, such gamma rays cannot travel more than $\sim 1$ Mpc
because of electron-positron pair creation with a radio background
\citep{pro96}.
The secondary emission via inverse Compton of cosmic microwave background photons
may be observed as GeV-TeV photons.
\citet{raz04b} concluded that the corresponding time delays are
in the range of 10-100 s and the duration timescale may be
much longer.
These timescales are much longer than the lifetime of $K^0_{\rm L}$ in the
observer's frame.
Therefore, unfortunately, the GeV-TeV secondary photons from kaons
may not be distinguishable from other sources, such as $\pi^0$-decay.
There is a possibility that failed GRBs occur near/in our
galaxy.
As \citet{mes01} pointed out, for very extended or slowly
rotating stars, the jet may be unable to break through the envelope of a
massive star. However, penetrating and choked jets will produce, by
photomeson interactions of accelerated protons, a burst of $\gtrsim 5$ TeV
neutrinos while propagating in the envelope \citep{razzaque04}.
Such neutrinos from nearby galaxies may be detectable by a $1 {\rm km}^2$ detector.
In failed GRBs, kaons may be also produced through the photomeson interactions
\citep{and05}.
When these classes of GRBs are taken into account, the chance probability
of detecting neutrinos will be enhanced considerably. 
In previous works, the neutrino
spectrum was estimated
assuming that the baryon density in the jet is sufficiently high
that the $pp$ process is the dominant process to produce neutrinos.
However, the efficiency depends sensitively
on the baryon density. So, there is a possibility
that photomeson production becomes the dominant process to
produce neutrinos in the failed GRBs.
In such a case, the spectrum of neutrinos will depend
sensitively on the photon spectrum. So, the detection of high-energy
neutrinos from such kaons will give us a clue to help understand
the physical conditions of failed GRBs, too.

\begin{acknowledgments}
First of all, we thank the referee for
his useful comments and advice.
We appreciate that S. Schadmand gave us information
on the cross sections of pion production.
We also thank T. Hyodo for his advice on the treatment
of kaon production processes.
KA acknowledges useful advice by S. Inoue.
This work is in part supported by a Grant-in-Aid for the 21st Century
COE ``Center for Diversity and Universality in Physics'' from the
Ministry of Education, Culture, Sports, Science and Technology of
Japan. S.N. is partially supported by Grants-in-Aid for Scientific
Research from the Ministry of Education, Culture, Sports, Science and
Technology of Japan through grants 14102004, 14079202, and 16740134.
\end{acknowledgments}

\end{document}